# Contrasting behavior of covalent and molecular carbon allotropes exposed to extreme ultraviolet and soft X-ray free-electron laser radiation

M. Toufarová[1,2], V. Hájková[1], J. Chalupský[1], T. Burian[1,3], J. Vacík[4], V. Vorlíček[1], L. Vyšín[1,2], J. Gaudin[5], N. Medvedev[1,6,*], B. Ziaja[7,8], M. Nagasono[9], M. Yabashi[9], R. Sobierajski[10], J. Krzywinski[11], H. Sinn[12], M. Störmer[13], K. Koláček[6], K. Tiedtke[14], S. Toleikis[14], L. Juha[1,6]

[1] Institute of Physics AS CR, v.v.i., Na Slovance 2, 182 21 Prague, Czech Republic

[2] Czech Technical University in Prague, Břehová 7, 115 19 Prague 1, Czech Republic

[3] Charles University in Prague, V Holešovičkách 2, 180 00 Prague 8, Czech Republic

[4] Nuclear Physics Institute AS CR, v.v.i., 250 68 Řež near Prague, Czech Republic

[5] Laboratoire CELIA, Université Bordeaux, 1 - 351 Cours de la Libération, 33405 Talence, France

[6] Institute of Plasma Physics AS CR, v.v.i., Za Slovankou 1782/3, 182 00 Prague, Czech Republic

[7] Center for Free Electron Laser Science at DESY, Notkestrasse 85, D-22603 Hamburg, Germany

[8] Institute of Nuclear Physics, Polish Academy of Sciences, Radzikowskiego 152, 31-342 Krakow, Poland

[9] RIKEN Harima Institute, The SPring-8 Center, Sayo-cho, Hyogo 679-5148, Japan

[10] Institute of Physics, Polish Academy of Sciences, Al. Lotników 32/46, PL-02-668 Warsaw, Poland

[11] LCLS at SLAC, 2575 Sand Hill Road, Menlo Park, CA 94025, USA

[12] European XFEL, Holzkoppel 4, D-22869 Schenefeld, Germany

[13] Helmholtz-Zentrum Geesthacht, Max-Planck-Strasse 1, D-21502 Geesthacht, Germany

[14] Deutsches Elektronen-Synchrotron - DESY, Notkestrasse 85, D-22603 Hamburg, Germany

## ABSTRACT

All carbon materials, e.g., amorphous carbon (a-C) coatings and $C_{60}$ fullerene thin films, play an important role in short wavelength free-electron laser (FEL) research motivated by FEL optics development and prospective nanotechnology applications. Responses of a-C and $C_{60}$ layers to the extreme ultraviolet (SCSS: SPring-8 Compact SASE Source in Japan) and soft X-ray (FLASH: Free-electron LASer in Hamburg) free-electron laser radiation are investigated by Raman spectroscopy, differential interference contrast and atomic force microscopy. A remarkable difference in the behavior of covalent (a-C) and molecular ($C_{60}$) carbonaceous solids is demonstrated under these irradiation conditions. Low thresholds for ablation of a fullerene crystal (estimated to be around 0.15 eV/atom for $C_{60}$ vs 0.9 eV/atom for a-C in terms of the absorbed dose) are caused by a low cohesive energy of fullerene crystals. An efficient mechanism of the removal of intact $C_{60}$ molecules from the irradiated crystal due to Coulomb repulsion of fullerene-cage cation radicals formed by the ionizing radiation is revealed by a detailed modeling.

* Corresponding author: nikita.medvedev@fzu.cz

# I. INTRODUCTION

The recent advent of short wavelength free-electron lasers (FEL) technology enabled a systematic investigation of changes occurring in materials irradiated by ultra-short pulses of extreme ultraviolet (30 nm < λ < 100 nm; XUV) and soft X-ray (0.3 nm < λ < 30 nm; SXR) coherent radiation [1,2]. Short wavelength radiation is absorbed due to an atomic photo-effect, depending mostly on the elemental composition and density of the sample [3]. At FEL facilities, carbonaceous materials are widely used as surface coatings covering optical elements developed for guiding and focusing FEL beams (amorphous carbon, a-C, is prevalent [4,5], but $C_{60}$ capping is also considered [6]). Such elements are heavily loaded by both thermal and radiation loads. Carbonaceous materials are also used as targets for imprinting FEL beams to reveal their characteristics (e.g. spatial distribution of radiation energy), test samples for nanopatterning induced by intense short-wavelength radiation etc. [7].

Interaction of solid a-C with short-wavelength FEL radiation has already been studied, while fullerenes have been exposed only as isolated $C_{60}$ molecules in a beam to X-ray FEL radiation at high fluences [8]. Not surprisingly, fullerene cages are usually decomposed into atomic ions under these severe irradiation conditions [8]. In this work, we demonstrate an unexpected behavior of a $C_{60}$ crystal under XUV and SXR irradiation, namely, a removal of intact $C_{60}$ molecules from the surface of the crystal. This conclusion is shown both by a detailed modeling as well as the supporting experimental data. Such $C_{60}$ removal occurs already at relatively mild irradiation fluences. In contrast, a-C targets can only be ablated at much higher doses, demonstrating the fundamental difference between van der Waals-bonded and covalently bonded materials.

## II. EXPERIMENTAL PART

In this study, 890 nm thick a-C and 200 nm thick $C_{60}$ layers were deposited on monocrystalline silicon substrates by magnetron sputtering and thermal evaporation, respectively. The deposition of fullerene molecules was carried out using specific deposition kinetics: the monocrystalline substrate was used because fullerenes cannot grow epitaxially on amorphous or polycrystalline materials. During the deposition, the substrate was kept at elevated temperature to induce desired surface mobility of the fullerene molecules. This was done to ensure a correct ordering of the fullerene thin films [9].

The attenuation lengths of FEL wavelengths of 13.5 nm [10] and 60 nm [11] (used in this experiment) in a fullerene layer are ~200 nm and ~10 nm, respectively, whereas in amorphous carbon they are ~160 nm and ~10 nm. As the attenuation lengths are typically small in comparison to the material thicknesses, we expect no effect of the finite sample size, except in the case of 13.5-nm-wavelength irradiation of $C_{60}$ crystal. This case will be discussed separately.

A surface roughness of a-C and $C_{60}$ samples was $1.02 \pm 0.07$ nm and $6.8 \pm 0.6$ nm correspondingly. The roughness of both pristine and irradiated surfaces was measured by AFM in the tapping mode. The measurements were carried out in the same way as the analysis reported in the earlier work [12].

In high vacuum interaction chambers, these materials were exposed to focused XUV laser radiation at the SCSS [13] (60 nm wavelength, see the scheme in Figure 1) and soft X-ray pulse at FLASH [14] (13.5 nm wavelength) large scale facilities. For a detailed description of the instrumentation and experimental procedures see Ref. [15]. An exposure was performed by single SCSS and FLASH pulses of 100 fs and 30 fs duration, respectively.

Differential interference contrast (DIC) microscopy, atomic-force microscopy (AFM) in tapping mode, and Raman spectroscopy excited by a 514.5-nm $Ar^+$ laser microbeam (MRS) were used to investigate damage patterns produced at various energy fluences adjusted by changing FEL pulse energy with gas attenuators and thin metallic foils. Pulse energy was always measured by photo-ionization Gas Monitor Detectors (GMD) filled with a suitable rare gas. The utilization of GMD was necessary, because FELs based on the SASE principle (Self-Amplified Spontaneous Emission [16,17]) provide pulse energies dramatically fluctuating shot to shot [1,2].

Series of single FEL shots were fired onto the samples fixed in the x-y-z micro-positioning system. An irradiation was conducted under normal incidence conditions. The area of the surface removed by ablation was determined by DIC and AFM techniques. For data analysis, the obtained values of eroded areas were plotted against the logarithm of FEL pulse energies. Assuming a stable beam profile, the low energy part of this plot was extrapolated to zero to obtain an ablation threshold energy; as an example, results for SCSS illuminated a-C are shown in Figure 2. Such a fluence scan method, with which the effective beam area and ablation threshold can be determined, is a standard technique used during damage experiments. Typically, few tens to hundreds of imprints are created to construct the Liu's plot and the derived f-scan curve, as described in details in Ref. [18]. This procedure was used for both the $C_{60}$ and a-C samples.

A dependence of the "peak-to-threshold" ratio (i.e., ablation threshold energy, $E_{TH}$, over the energy of a particular FEL pulse) on a corresponding eroded area for SCSS irradiated a-C and $C_{60}$ is shown in Figure 3. Such a plot is usually called an f-scan. The area under the curve is equal to the effective beam area $A_{EFF}$ which enables evaluation of the fluence threshold for the investigated material as follows:

$$F_{TH} = \frac{E_{TH}}{A_{EFF}} \tag{1}$$

A normalized sum of two exponential functions was used for fitting the experimental datasets. As it was proved by previous experiments, this function describes a typical FEL beam in terms of a superposition of a narrow intense central peak and wide wings. It follows from Figure 3 that the determination of an effective beam area is insensitive to irradiated material, at least for a particular interaction experiment. The f-scans exhibit the same trend, and fitting functions overlap with one another; they provide almost the same values of the FEL beam effective area even if some parts of the f-scans related to the smallest area are missing (i.e., low FEL energies in heavily attenuated pulses, which are difficult to measure accurately because of the GMD noise and limited sensitivity).

### III. THEORY AND COMPUTER SIMULATIONS

In order to model a behavior of carbonaceous materials under a femtosecond FEL irradiation, the recently developed hybrid code XTANT was extended correspondingly and applied [19,20]. The code combines a few theoretical schemes into one model to describe important aspects of XUV or x-ray radiation effects in matter:

(i) FEL irradiation, photo absorption and excitation of electrons, as well as their secondary cascading, are described within an event-by-event Monte Carlo (MC) framework. MC module traces all photo-electrons and impact-ionized electrons with energies above 10 eV.

(ii) Electrons at the bottom of the conduction band and in the valence band of the material are traced within a simplified Boltzmann kinetic approach assuming Fermi-Dirac distribution. The kinetic equation includes energy and particle source and sink terms, responsible for energy exchange with the highly excited electrons (traced within the MC module) and with the atomic lattice [19,20]. These electrons populate the energy levels corresponding to the transient band structure of the irradiated sample.

(iii) Transient band structure, as well as the forces acting on atoms, are obtained by a diagonalization of the transient Hamiltonian within the transferable tight binding (TB) method. All the details of TB parameterization can be found in [21]. The potential energy surface, describing short-range covalent bonds, depends on the transient state of all the atoms within the simulation box and on the electron distribution function.

Although XTANT explicitly accounts for the band structure of the molecules in the simulations, the charge transfer is not explicitly modeled. Instead, as mentioned above, high-energy electrons are treated within the event-by-event Monte Carlo scheme. These electrons are quickly spread among the C60 molecules, producing a nearly homogeneous distribution of the excited electrons. Part of these electrons is removed from the simulation box due to emission from the surface, as described in more detail below. Then, the unbalanced charge left in the system is distributed among all the atoms homogeneously (fractional charge), as the statistical average from the Monte Carlo simulation. Thus, two channels of excitation of $C_{60}$ cages from the ground state are included in our model: (a) heating of the electronic subsystem due to interaction with high-energy electrons (change of the temperature and the chemical potential of the corresponding Fermi-Dirac function), and (b) unbalanced charge left after electron emission from the surface.

(iv) Atomic motion is traced within classical molecular dynamics (MD) scheme on the potential energy surface calculated within the above-mentioned TB approximation. In addition, the energy exchange between the atoms and the low-energy electrons from the valence and the conduction band is calculated via Boltzmann electron-ion (electron-phonon) collision integral [20].

In order to account for peculiarities of the studied problem, additional modules had to be introduced into XTANT. First, van der Waals forces were included in addition to the covalent bonds to describe the long-range binding between separate $C_{60}$ cages, which also plays a role for a-C (and graphite). This was done by adding the classical long-range potential within Girifalco's model based on Lenard-Jones (6,12) potential, cut at short and large distances (similar to Refs. [22,23]):

$$V_{vdW}(r) = \begin{cases} ar^5 + br^4 + cr^3 + dr^2 + er + f, \text{ for } 2.5 < r < 3.4 \text{ Å} \\ \frac{C_{12}}{r^{12}} + \frac{C_6}{r^6}, \text{ for } 3.4 \text{ Å} < r < 5 \text{ Å} \\ a_L r^3 + b_L r^2 + c_L r + d_L, \text{ for } 5 \text{ Å} < r < 6 \text{ Å} \end{cases}, \quad (2)$$

Here $V_{vdW}(r)$ is the total van der Waals (vdW) potential. At short distances it is replaced by a polynomial fitted to smoothly join the exact potential at the point of 3.4 Å (the function itself as well as its first and second derivatives), and smoothly approaching zero at 2.5 Å. More details on the potential and the coefficients used are given in Appendix A.

The potential uses soft cut-offs to avoid an overlap with the short-range covalent bonds that are described within TB, and at large distance, to treat the long-range force. It is also assumed to be not explicitly dependent on the state of the electronic system; instead, an unbalanced charge due to electron emission is treated separately, see below.

Second, as XTANT uses the periodic boundary conditions, no direct photoemission was included in the original code. To account for the electron emission from the surface, the following scheme is proposed here. A highly excited electron is artificially removed from the supercell ('emitted') after a certain number of secondary collisions performed. The number of collisions before the emission is estimated as an average ratio between a photo-electron range and its inelastic mean free path [24]. It depends on the initial photo-electron energy, and, thus, on the photon energy in the pulse. For a FLASH pulse, it is estimated to be 4-5 collisions on average, while for SCSS case, it is ~2 collisions since the photon attenuation in the latter case is much shorter. This means that electrons (excited closer to a surface) can be emitted faster. The electron ranges are significantly shorter than the photon attenuation lengths thus electrons are only emitted from a near-surface region, while their effects of the back surface and the substrate are neglected.

This scheme allows accounting for secondary electron cascades, simultaneously producing an unbalanced charge in the system after an electron removal. After an electron is emitted (removed from the system), the corresponding positive charge left in the system is distributed equally among all the atoms in the simulation box assuming instant charge transfer. This follows the scheme of the instant thermalization and the Ehrenfest dynamics framework assumed in the model [19,20]. Since the accumulated charge is only a fraction of percent (as will be shown below), we neglect possible effects of charge inhomogeneity; such an approximation should not significantly influence an outcome of the simulation.

The unbalanced charge accumulated after an electron emission then contributes to the interatomic forces. The classical Coulomb potential is added within the MD part to trace its effect on the atomic motion. The Coulomb potential is softly cut off at a large distance covering the interaction of all atoms inside of the simulation box as follows (in SI units):

$$V_C(r) = \frac{Q_1 Q_2}{4\pi\varepsilon_0 r} \frac{1}{1 + exp((r - r_c)/d_c)}, \qquad (3)$$

Here $Q_1$ and $Q_2$ are the atomic charges, corresponding to the unbalanced charge built-up in the system. The cut-off multiplier (a Fermi-like cut-off function) has the following parameters $d_c = 0.1$ Å, and $r_c$ which must exceed the maximal size of the simulated sample (thus, the cut off distance depends on the number of atoms and their geometry modeled; e.g. $r_c = 30$ Å for three $C_{60}$ cages modeled). The details of the potential and parameter analysis are presented in the Appendix B.

Calculations were performed for the FEL parameters corresponding to the experiments: photon energy was 92 eV for FLASH and 21 eV for SCSS case; pulse durations were 30 fs and 100 fs, correspondingly.

Typical simulation supercell contained 120 or 180 atoms inside, i.e. two or three $C_{60}$ cages were explicitly modeled. They were separated from each other by a distance of 3.35 Å corresponding to the equilibrium position of the van der Waals potential for simple cubic (scc) $C_{60}$. Periodic boundary conditions along X and Y axes were used, separating the image molecules by the same inter-cage distance of 3.35 Å. The system then represents a molecular crystal with free boundaries along Z axis allowing for the material removal from the crystal.

It is known from the literature that a solid $C_{60}$ crystal has two stable phases: simple cubic crystal (scc), and face-centered cubic phase (fcc). It was shown both, experimentally and theoretically, that scc phase is stable at temperatures below 250-260 K [25], whereas fcc phase is stable at room temperatures [26]. We performed simulations for both phases.

In case of amorphous carbon, the sample contained 216 atoms. The initial amorphous state was prepared by quenching of the equilibrated melted state. The initial state preparation also included extensive cross-checks to confirm that there are no artificial pieces of diamond/graphite or voids left in the homogeneous sample. After that, an additional thermalization at the room temperature prior to an FEL pulse was performed ensuring stability of the sample. Parrinello-Rahman MD scheme was used to model the a-C, accounting for changes in the volume of the super-cell and eventual material ablation [19]. A set of simulation runs with different initial conditions (random velocities of atoms) was performed to confirm reproducibility of the damage threshold.

Note that in case of SCSS photon energy (21 eV), the contribution of direct photoabsorption by the free electrons (inverse Bremsstrahlung) might be not negligible [27]. This process was not taken into account, which might slightly influence the calculation results for the SCSS case.

## IV. RESULTS AND DISCUSSION

### 1. Experimental resuls

Single-shot ablation thresholds determined in experiment by applying the above mentioned procedure (for more details see Ref. [7]) are summarized in Table 1. Damage threshold fluences for fullerene exposed to SCSS and FLASH radiation are 4-fold and 9-fold lower, respectively, than in a-C irradiated at the same wavelength and pulse duration. This difference follows from the fact that a-C represents covalent carbonaceous solids while $C_{60}$ fullerene is a typical example of a molecular solid, in where $C_{60}$ clusters (i.e., fullerene cages) are only bound together by a weak intermolecular interaction, van der Waals forces. Thus, a significantly lower energy density has to be achieved to evaporate $C_{60}$ cages from the sample surface in vacuum than for a-C.

To calculate the mean energy density at the ablation threshold from the experimental fluence the following equation was used:

$$D_{TH} = \frac{F_{TH}}{l_{ATT}} \qquad (4)$$

Threshold energy densities $D_{TH}$ were determined from the threshold fluences, $F_{TH}$ from Eq.(1), obtained experimentally and the attenuation lengths $l_{ATT}$ of ~200 nm (FLASH [10]) and ~10 nm (SCSS [11]), see Table 1. Divided by the molecular density of fullerene cages (1.4 x $10^{21}$ $cm^{-3}$), it indicates that 0.1 and 5.0 photo-ionization events per one $C_{60}$ cage are required to initiate fullerene ablation by 13.5 nm and 60 nm-wavelength FEL radiation, respectively. Those values correspond to 0.15 eV/atom (FLASH) and between 0.44 and 1.1 eV/atom (SCSS; the uncertainty appears due to the uncertainty in the photon attenuation length).

This difference can be expected, as a photo-electron liberated from a fullerene cage by 92 eV radiation (i.e. FLASH tuned at 13.5 nm) carries enough energy to trigger a collisional ionization cascade resulting in formation of several fullerene cation radicals in the close neighborhood. In contrast, a photo-electron following the interaction of a single 21 eV photon (i.e., SCSS tuned at 60 nm) can ionize only one additional cage. So, more XUV photons should be absorbed in a volume unit to achieve the same total ionization density as SXR irradiation does.

In Figure 4, no change in the positions of the peaks and no new peaks appearing in Raman spectra of $C_{60}$ can be seen, although the irradiated material has expanded. This means that the expansion cannot be attributed to graphitization (as also supported by modeling results below), which is typical for irradiated a-C (see, for example, Figure 5, and more details in Appendix C). Fullerene expansion at higher fluences can be explained by FEL-induced damage on the substrate resulting in the increase in the damage pattern's outer contour. At a lower FEL fluence and longer wavelengths (the attenuation length of 60 nm-wavelength radiation in solid $C_{60}$ is of the order of 10 nm which is more than an order of magnitude smaller than the sample thickness) the silicon carbon interface cannot be directly influenced by the deposited FEL energy, only its erosion has been registered by AFM in the FEL-irradiated $C_{60}$ material (Figure 6a). Thus, the effect of the substrate on the carbon target can be considered negligible.

Raman spectra (Figure 6b) do not indicate any sign of amorphous and/or graphitic carbon formation in the irradiated area (Figure 6c). Formation of highly disordered nanographite, if occurred, would give a clear contribution to the signal (see Appendix C). We deduce from this fact that the decomposition of $C_{60}$ does not occur, and $C_{60}$ molecules are removed from the surface intact as molecular ions (this possibility will be tested below by modeling). The AFM reveals a slightly increased roughness in the ablated area, i.e., from $6.8 \pm 0.6$ nm to $9.2 \pm 0.8$ nm, but does not show any change in the $C_{60}$ surface flatness that could indicate a change in the substrate shape.

Since fullerenes can be efficiently transformed into other covalently bound carbon phases, including graphite and amorphous carbon, by different means: at elevated temperatures [28] and/or pressures, when irradiated with conventional long-wavelength lasers [29] (and Appendix C), and exposed to electromagnetic [30] and particle [31] ionizing radiation, the absence of such changes in FEL-induced damage patterns indicates the above described non-thermal mechanisms of fullerene erosion, supporting the modeled results.

The above described fullerene removal mechanism is supported by investigation of the surface morphology and chemical constitution changes by AFM and Raman spectroscopy. Both material removal (erosion, i.e., desorption and ablation) and expansion (extrusion) were observed in $C_{60}$ and a-C materials irradiated by SCSS and FLASH ultra-short laser pulses. The expansion occurred more frequently in irradiated a-C than in $C_{60}$-crystal. For a-C the damage threshold was found to be around 170 mJ/cm$^{-2}$ for 13.5 nm FLASH radiation [32]; the expansion is preceding ablation, for which the damage threshold corresponds to ~0.88 eV/atom in terms of the absorbed dose.

Raman spectra also do not show any sign of the substrate-induced processes. It agrees with the above-made estimate based on the photon attenuation lengths being typically smaller than the sample thickness. Only in the case of the FLASH shining on 200-nm thick $C_{60}$-crystal, there might be some effects induced by the part of the pulse penetrating into the substrate. However, electrons emitted from the substrate have short ranges and do not reach the front-surface, thereby not affecting observable results. The back surface seems to be unaffected either, as in this case one would expect to observe effects similar to Ref. [30], which were not observed in the current experiment.

We also did not observe any significant large-scale modifications of the surface of the irradiated targets, as the surface in the damage pattern was smooth. Although laser-induced periodic surface structures (LIPSS) with a spatial period related to the FEL wavelength were reported for a very small fraction of the 98 nm FEL irradiated a-C coatings [4], neither a-C nor $C_{60}$ surfaces exposed to a SCSS/FLASH single shot exhibit any formation of such periodic structures here. This is very likely caused by a lack of back reflectivity of XUV/SXR radiation from smooth sample surfaces illuminated under normal incidence conditions. An absence of the spontaneously created ripples on FEL illuminated surfaces could be useful for imprinting a de-magnified pattern from a mask [33] and/or an interferometer [34].

## 2. Simulated damage process

Our model results demonstrate that irradiated $C_{60}$ crystal disintegrates into single intact fullerenes. The observed fullerene behavior is caused by a Coulomb explosion induced by charging fullerene cages. This unbalanced charge is produced due to their photo-ionization by XUV/SXR laser radiation and their impact ionization by photo-electrons and secondary electrons. When the energy of an excited electron is above the work function of $C_{60}$ (which is only 7.6 eV), the electron can be emitted leaving a positive charge behind. The repulsive forces between neighboring fullerene cation radicals then decompose the molecular crystal structure, releasing fullerene cages into the vacuum.

Figure 7 shows calculated snapshots of the system at different times following the FLASH irradiation of $C_{60}$ crystal. This figure demonstrates an example of scc structure; the fcc structure simulation looks similar and its damage threshold is lower only by ~10%. This could be expected from the considerations of the cohesive energy: for the fcc and for the scc phases it differs only by ~10% (E=-1.772 eV/$C_{60}$ vs E=-1.968 eV/$C_{60}$, respectively, or in the absolute value only by 0.0033 eV/atom [26]). Below we provide the estimations for both of the structures.

In particular, Figure 7 shows: (i) the sub-threshold dose for which the crystal stays intact, (ii) the above-threshold dose showing a intermolecular Coulomb explosion, and (iii) an 'artificial' model case excluding electron emission from the $C_{60}$ crystal. In the latter case no ablation takes place even for the above-threshold dose. This clearly demonstrates the Coulomb explosion as the mechanism for $C_{60}$ removal from the crystal. Note that the $C_{60}$ molecules removed stay intact without breaking apart into smaller atomic fragments. These results support the scenario of $C_{60}$-crystal damage inferred from the experimental data above.

The damage threshold dose is estimated as the threshold charge built-up, at which the Coulomb repulsive potential overcomes the attractive van der Waals potential. As the long-range force, the Coulomb potential contributes from all the affected $C_{60}$ cages. Considering the excited electron range evaluated from its loss function, after the FLASH irradiation on average six near-surface layers of $C_{60}$ molecules in the crystal can have a significant unbalanced charge due to an electron emission; in case of SCSS irradiation it is only two layers. Summing up the potential from the corresponding number of layers results in the threshold charge of 0.0018 electrons/atom for fcc structure (0.002 for scc) for FLASH case, and 0.0057 for fcc (0.0064 for scc $C_{60}$) electrons/atom for SCSS case. They would be needed to initiate the intermolecular Coulomb explosion. Note that these charges are only a fraction of atomic density, confirming an assumption of a small charge inbuilt made above. XTANT calculations demonstrate that the respective charges are reached for the absorbed doses of ~0.18 eV/atom for fcc in FLASH case (0.21 eV/atom for scc) and 0.67 eV/atom for fcc (0.75 eV/atom for scc) in SCSS case, which are close to the experimentally observed damage thresholds (0.15 eV/atom for FLASH and 0.44-1.1 eV/atom for SCSS).

The intermolecular Coulomb explosion takes place on a timescale of over a picosecond due to a large inertia of $C_{60}$ cages, although the unbalanced charge is established within the FEL pulse duration (sub-100 fs). At the considered irradiation conditions of FLASH and SCSS, the electron cascades are extremely fast, finishing within a few femtoseconds [24], after which there is no additional electron emission that would increase the ionic charge.

For the under-threshold absorbed dose, $C_{60}$ crystal is expanding due to the presence of the repulsive Coulomb field. The van der Waals bonds between $C_{60}$ cages in the left panel of Figure 7 are elongated indicating below-threshold expansion of the sample (in agreement with the experiment above); whereas without accounting for the Coulomb field due to the unbalanced charge, no such elongation is observed in the simulation (right panel of the same figure).

The results of irradiation of amorphous carbon are demonstrated in Figure 8. Calculations with XTANT for a-C showed that for the considered parameters of FLASH irradiation, the damage threshold is ~0.85-0.9 eV/atom (cf. experimental dose of 0.88 eV/atom, see above). Here, the damage threshold is defined as a dose needed to initiate ablation/disintegration of the amorphous sample. For an above-threshold dose, the irradiated sample breaks apart into molecular fragments, and the volume of the modeled supercell expands further. Both effects can be observed in the simulation and distinguished from the below-damage case where no ablation was observed.

For the below-threshold absorbed dose (< 0.85 eV/atom), Figure 8 shows expansion of the irradiated material, which saturates after ~2.5 ps (see Appendix D also showing an analysis of the volume of the modeled super-cell). As atomic snapshots in Figure 8 indicate, it proceeds similar to a graphitization, although the formed graphite-like planes are bent and defected. This below-threshold expansion reproduces the experimental finding described in the previous section.

Unfortunately, XTANT is unable to model irradiation of a-C with the SCSS pulse, due to a very large gradient of the photo-absorption profile (since the photon attenuation length is on the order of 10 nm). Such strong gradients violate the assumption of periodic boundaries used in XTANT for the a-C case, and ultrafast particle and heat transport in the electronic system must be included for a meaningful comparison with experimental data. Such a study will be a topic of a separate dedicated work.

## CONCLUSIONS

In conclusion, the low damage thresholds, non-thermal character of material erosion (via intermolecular Coulomb explosion), and the lack of chemical and/or phase transformations in FEL-irradiated areas suggest $C_{60}$ fullerene layers as a promising material for efficient and clean surface nanopatterning induced by short-wavelength lasers. This damage mechanism is described in detail by our model which results support the experimental findings.

In contrast, amorphous carbon is more resistant to an FEL radiation due to the strength of its covalent bonds. Under irradiation, a-C undergoes partial graphitization and material expansion, as also supported by modeling results. At even higher fluences, material ablation occurs.

## ACKNOWLEDGEMENTS

This work was supported by the Czech Science Foundation (Grant No. 14-29772S) and by the Czech Ministry of Education, Youth and Sports (Grants No. LM2015083, No. CZ.02.1.01/0.0/0.0/16_013/0001552, and No. LG15013). B.Z. thanks Zoltan Jurek for discussions.

## APPENDIX A: VAN DER WAALS POTENTIAL

Van der Waals potential is assumed within Girifalco's model with Lenard-Jones (6,12) potential [22,23], with an additional soft cut-offs at small and large distances, Eq.(2). We reproduce this equation here again for convenience of the reader:

$$V_{vdW}(r) = \begin{cases} ar^5 + br^4 + cr^3 + dr^2 + er + f, \text{ for } 2.5 \text{ Å} < r < 3.4 \text{ Å} \\ \frac{C_{12}}{r^{12}} + \frac{C_6}{r^6}, \text{ for } 3.4 \text{ Å} < r < 5 \text{ Å} \\ a_L r^3 + b_L r^2 + c_L r + d_L, \text{ for } 5 \text{ Å} < r < 6 \text{ Å} \end{cases}, \quad \text{(A1)}$$

Here $V_{vdW}(r)$ is the total van der Waals (vdW) potential. At short distances it is replaced by a polynomial fitted to smoothly join the exact potential at the point of 3.4 Å (the function itself as well as its first and second derivatives), and smoothly approaching zero at 2.5 Å (fifth order polynomial ensures that the potential approaches zero at short distances from the positive side, 0+). Thereby, it does not overlap with the short-range covalent bonds described within TB scheme. At large distances it is replaced by another polynomial to match the vdW potential (and its first derivative) at the distance of 5 Å, while turning to zero at 6 Å. At distances $r < 2.5$ Å or $r > 6$ Å the potential is set to zero, $V_{vdW}(r) = 0$. The distance 6 Å is chosen as an intermediate distance between a second-nearest graphene planes in graphite, thus including only the interaction between the nearest neighbor planes. The coefficients of the potential and polynomials are listed in Table II. Note that these coefficients are fitted only to reproduce the correct minimum of the van der Waals potential, and its qualitative shape, but might not precisely reproduce the vibrational frequencies as they are not the topic of the current investigation.

## APPENDIX B: COULOMB POTENTIAL AND UNBALANCED CHARGE

Coulomb potential, $V_c(r)$, is introduced with the soft cut-off in the following way, Eq.(3), which we also reproduce here for convenience:

$$V_C(r) = \frac{Q_1 Q_2}{4\pi\varepsilon_0 r} \frac{1}{1 + exp((r - r_c)/d_c)}, \quad \text{(A2)}$$

here $Q_1$ and $Q_2$ are the atomic charges, corresponding to the unbalanced charge built-up in the system; the Fermi-like cut-off function has the following parameters: $d_c = 0.1$ Å, and $r_c$, which must exceed the maximal size of the simulated sample (e.g. $r_c = 30$ Å for three $C_{60}$ cages modeled).

The influence of the chosen number of collisions, after which an electron is considered to be emitted, was analyzed. An emitted electron is removed from the MC part of the model. The number of these electrons is counted. The corresponding positive charge left in the system is distributed equally among all the atoms in all the $C_{60}$ molecules, creating Coulomb repulsive potential. Figure 9 shows that the largest unbalanced charge is built-up in the case when an electron is considered to be emitted after one collision. This value is decreasing with increase of the number of collisions. In case if more collisions are allowed, electrons have a higher chance to lose their energy and to fall back to the bottom of the conduction band – below the work function, from where they can no longer be emitted. This, however, depends on the photon energy: e.g., it would require many more collisions for electrons produced by hard X-ray photons. In the calculations, the number of collisions is chosen from the considerations of an electron mean free path and its total range. For FLASH case, it is estimated to be 4 collisions, whereas for SCSS case it is 2 collisions, which occur within a distance of a few nm, much smaller than the sample size.

### APPENDIX C: PROLONGED EXPOSURE OF $C_{60}$ CRYSTAL TO VISIBLE LIGHT

Prolonged exposure of $C_{60}$ crystal to visible light irradiation (cw mode, 514.5 nm wavelength) during Raman spectra collection induces noticeable changes in the material, as shown in Figure 10. These Raman spectra of initially pristine crystal show eventual decomposition and amorphization. Each spectrum is taken two minutes after the previous one. The fullerene exposure to cw 514.5-nm laser radiation clearly shows the formation of amorphous/graphitic phases. Thus, one can conclude that damage of individual $C_{60}$ cages is well noticeable in Raman spectra.

Although total doses in the single-shot FEL exposure are comparable to the above mentioned cw-Vis laser irradiation, we do not see such a behavior in the SXR/XUV-FEL case. The D/G-modes change is clearly visible in a-C spectra (as was already reported in [35]), while fullerene layers do not exhibit such a behavior. No significant fraction of small carbonaceous species was found re-deposited in the crater and/or on its rim. Absence of any traces of amorphized or decomposed carbon after XUV/SXR FEL radiation reported in the main text supports the hypothesis that intact $C_{60}$ molecules are emitted.

### APPENDIX D: COMPUTATIONAL EVALUATION OF DAMAGE THRESHOLD

Damage threshold for $C_{60}$ crystal is defined by the unbalanced-charge built at which the induced Coulomb field exceeds the van der Waals potential keeping $C_{60}$ cages together. This depends on two factors: (i) energy deposited into the sample by an FEL pulse, which defines the number of excited electrons that can potentially be emitted; and (ii) a probability of an electron emission. Electrons are emitted from the surface only if their energy remaining after cascading is sufficiently high to overcome the work function of the material.

The threshold charge is shown in Figure 11. It is defined as the charge for which Coulomb potential overcomes the vdW potential, making the total potential energy in the system positive, and triggering disintegration of the $C_{60}$ crystal. Knowing the threshold charge, we can then run a set of XTANT simulations to evaluate which absorbed dose produces a sufficient electron emission.

In the case of amorphous carbon, the damage is defined as the ablation threshold, which we can detect by the volume expansion of the simulated supercell within the Parrinello-Rahman MD scheme. Figure 12 shows the threshold is at ~0.9 eV/atom; for lower absorbed doses the supercell volume expands up to the time of ~2.5 ps, after which the expansion stops without ablation. For higher doses, the volume expands further, and sample disintegrates into fragments (see the main text).

## REFERENCES

[1]     E. L. Saldin, E. A. Schneidmiller, and M. V. Yurkov, *The Physics of Free Electron Lasers* (Springer Berlin Heidelberg, Berlin, Heidelberg, 2000).

[2]     P. Schmüser, M. Dohlus, J. Rossbach, and C. Behrens, *Free-Electron Lasers in the Ultraviolet and X-Ray Regime* (Springer International Publishing, Cham, 2014).

[3]     D. Attwood, *Soft X-Rays and Extreme Ultraviolet Radiation: Principles and Applications* (Cambridge University Press, Cambridge,1999).

[4]     B. Steeg, L. Juha, J. Feldhaus, S. Jacobi, R. Sobierajski, C. Michaelsen, A. Andrejczuk, and J. Krzywinski, Applied Physics Letters **84**, 657 (2004).

[5]     M. Störmer, F. Siewert, J. Buchheim, A. Pilz, M. Kuhlmann, E. Ploenjes, and K. Tiedtke, in *Proc. SPIE 9207, Advances in X-Ray/EUV Optics and Components IX*, edited by C. Morawe, A. M. Khounsary, and S. Goto (International Society for Optics and Photonics, 2014), p. 92070H.

[6]     J. A. Méndez, J. I. Larruquert, and J. A. Aznárez, Applied Optics **39**, 149 (2000).

[7]     J. Chalupský, J. Krzywinski, L. Juha, V. Hájková, J. Cihelka, T. Burian, L. Vyšín, J. Gaudin, A. Gleeson, M. Jurek, A. R. Khorsand, D. Klinger, H. Wabnitz, R. Sobierajski, M. Störmer, K. Tiedtke, and S. Toleikis, Optics Express **18**, 27836 (2010).

[8]     B. F. Murphy, T. Osipov, Z. Jurek, L. Fang, S.-K. Son, M. Mucke, J. H. D. Eland, V. Zhaunerchyk, R. Feifel, L. Avaldi, P. Bolognesi, C. Bostedt, J. D. Bozek, J. Grilj, M. Guehr, L. J. Frasinski, J. Glownia, D. T. Ha, K. Hoffmann, E. Kukk, B. K. McFarland, C. Miron, E. Sistrunk, R. J. Squibb, K. Ueda, R. Santra, and N. Berrah, Nature Communications **5**, (2014).

[9]     K. M. Kadish and R. S. Ruoff, *Fullerenes : Chemistry, Physics, and Technology* (Wiley-Interscience, 2000).

[10]    B. L. Henke, E. M. Gullikson, and J. C. Davis, Atomic Data and Nuclear Data Tables **54**, 181

(1993).

[11]   E. D. Palik, *Handbook of Optical Constants of Solids* (Academic Press, San Diego, 1985).

[12]   J. Chalupský, L. Juha, V. Hájková, J. Cihelka, L. Vyšín, J. Gautier, J. Hajdu, S. P. Hau-Riege, M. Jurek, J. Krzywinski, R. A. London, E. Papalazarou, J. B. Pelka, G. Rey, S. Sebban, R. Sobierajski, N. Stojanovic, K. Tiedtke, S. Toleikis, T. Tschentscher, C. Valentin, H. Wabnitz, and P. Zeitoun, Optics Express **17**, 208 (2009).

[13]   T. Shintake, H. Tanaka, T. Hara, T. Tanaka, K. Togawa, M. Yabashi, Y. Otake, Y. Asano, T. Bizen, T. Fukui, S. Goto, A. Higashiya, T. Hirono, N. Hosoda, T. Inagaki, S. Inoue, M. Ishii, Y. Kim, H. Kimura, M. Kitamura, T. Kobayashi, H. Maesaka, T. Masuda, S. Matsui, T. Matsushita, X. Maréchal, M. Nagasono, H. Ohashi, T. Ohata, T. Ohshima, K. Onoe, K. Shirasawa, T. Takagi, S. Takahashi, M. Takeuchi, K. Tamasaku, R. Tanaka, Y. Tanaka, T. Tanikawa, T. Togashi, S. Wu, A. Yamashita, K. Yanagida, C. Zhang, H. Kitamura, and T. Ishikawa, Nature Photonics **2**, 555 (2008).

[14]   W. Ackermann, G. Asova, V. Ayvazyan, A. Azima, N. Baboi, J. Bähr, V. Balandin, B. Beutner, A. Brandt, A. Bolzmann, R. Brinkmann, O. I. Brovko, M. Castellano, P. Castro, L. Catani, E. Chiadroni, S. Choroba, A. Cianchi, J. T. Costello, D. Cubaynes, J. Dardis, W. Decking, H. Delsim-Hashemi, A. Delserieys, G. Di Pirro, M. Dohlus, S. Düsterer, A. Eckhardt, H. T. Edwards, B. Faatz, J. Feldhaus, K. Flöttmann, J. Frisch, L. Fröhlich, T. Garvey, U. Gensch, C. Gerth, M. Görler, N. Golubeva, H.-J. Grabosch, M. Grecki, O. Grimm, K. Hacker, U. Hahn, J. H. Han, K. Honkavaara, T. Hott, M. Hüning, Y. Ivanisenko, E. Jaeschke, W. Jalmuzna, T. Jezynski, R. Kammering, V. Katalev, K. Kavanagh, E. T. Kennedy, S. Khodyachykh, K. Klose, V. Kocharyan, M. Körfer, M. Kollewe, W. Koprek, S. Korepanov, D. Kostin, M. Krassilnikov, G. Kube, M. Kuhlmann, C. L. S. Lewis, L. Lilje, T. Limberg, D. Lipka, F. Löhl, H. Luna, M. Luong, M. Martins, M. Meyer, P. Michelato, V. Miltchev, W. D.

Möller, L. Monaco, W. F. O. Müller, O. Napieralski, O. Napoly, P. Nicolosi, D. Nölle, T. Nuñez, A. Oppelt, C. Pagani, R. Paparella, N. Pchalek, J. Pedregosa-Gutierrez, B. Petersen, B. Petrosyan, G. Petrosyan, L. Petrosyan, J. Pflüger, E. Plönjes, L. Poletto, K. Pozniak, E. Prat, D. Proch, P. Pucyk, P. Radcliffe, H. Redlin, K. Rehlich, M. Richter, M. Roehrs, J. Roensch, R. Romaniuk, M. Ross, J. Rossbach, V. Rybnikov, M. Sachwitz, E. L. Saldin, W. Sandner, H. Schlarb, B. Schmidt, M. Schmitz, P. Schmüser, J. R. Schneider, E. A. Schneidmiller, S. Schnepp, S. Schreiber, M. Seidel, D. Sertore, A. V. Shabunov, C. Simon, S. Simrock, E. Sombrowski, A. A. Sorokin, P. Spanknebel, R. Spesyvtsev, L. Staykov, B. Steffen, F. Stephan, F. Stulle, H. Thom, K. Tiedtke, M. Tischer, S. Toleikis, R. Treusch, D. Trines, I. Tsakov, E. Vogel, T. Weiland, H. Weise, M. Wellhöfer, M. Wendt, I. Will, A. Winter, K. Wittenburg, W. Wurth, P. Yeates, M. V. Yurkov, I. Zagorodnov, and K. Zapfe, Nature Photonics **1**, 336 (2007).

[15]  R. Sobierajski, M. Jurek, J. Chalupský, J. Krzywinski, T. Burian, S. D. Farahani, V. Hájková, M. Harmand, L. Juha, D. Klinger, R. A. Loch, C. Ozkan, J. B. Pełka, K. Sokolowski-Tinten, H. Sinn, S. Toleikis, K. Tiedtke, T. Tschentscher, H. Wabnitz, and J. Gaudin, Journal of Instrumentation **8**, P02010 (2013).

[16]  S. V. Milton, E. Gluskin, N. D. Arnold, C. Benson, W. Berg, S. G. Biedron, M. Borland, Y.-C. Chae, R. J. Dejus, P. K. Den Hartog, B. Deriy, M. Erdmann, Y. I. Eidelman, M. W. Hahne, Z. Huang, K.-J. Kim, J. W. Lewellen, Y. Li, A. H. Lumpkin, O. Makarov, E. R. Moog, A. Nassiri, V. Sajaev, R. Soliday, B. J. Tieman, E. M. Trakhtenberg, G. Travish, I. B. Vasserman, N. A. Vinokurov, X. J. Wang, G. Wiemerslage, and B. X. Yang, Science **292**, (2001).

[17]  Z. Huang and K.-J. Kim, Physical Review Special Topics - Accelerators and Beams **10**, 34801 (2007).

[18]  J. Chalupský, T. Burian, V. Hájková, L. Juha, T. Polcar, J. Gaudin, M. Nagasono, R.

Sobierajski, M. Yabashi, and J. Krzywinski, Optics Express **21**, 26363 (2013).

[19]   N. Medvedev, H. O. Jeschke, and B. Ziaja, New Journal of Physics **15**, 15016 (2013).

[20]   N. Medvedev, Z. Li, V. Tkachenko, and B. Ziaja, Physical Review B **95**, 14309 (2017).

[21]   C. H. Xu, C. Z. Wang, C. T. Chan, and K. M. Ho, Journal of Physics: Condensed Matter **4**, 6047 (1992).

[22]   A. Carlson, An Extended Tight-Binding Approach for Modeling Supramolecular Interactions of Carbon Nanotubes, UNIVERSITY OF MINNESOTA, 2006.

[23]   A. Carlson and T. Dumitrică, Nanotechnology **18**, 65706 (2007).

[24]   N. Medvedev, Applied Physics B **118**, 417 (2015).

[25]   W. I. F. David, R. M. Ibberson, J. C. Matthewman, K. Prassides, T. J. S. Dennis, J. P. Hare, H. W. Kroto, R. Taylor, and D. R. M. Walton, Nature **353**, 147 (1991).

[26]   J. P. Lu, X.-P. Li, and R. M. Martin, Physical Review Letters **68**, 1551 (1992).

[27]   F. Wang, E. Weckert, and B. Ziaja, Journal of Plasma Physics **75**, 289 (2009).

[28]   M. R. Stetzer, P. A. Heiney, J. E. Fischer, and A. R. McGhie, Physical Review B **55**, 127 (1997).

[29]   H. M. Phillips, D. Sarkar, N. J. Halas, R. H. Hauge, and R. Sauerbrey, Applied Physics A Solids and Surfaces **57**, 105 (1993).

[30]   H. Klesper, R. Baumann, J. Bargon, J. Hormes, H. Zumaque-Diaz, and G. A. Kohring, Applied Physics A **80**, 1469 (2005).

[31]   T. Braun, H. Rausch, and J. Mink, Carbon **43**, 870 (2005).

[32]   J. Chalupský, V. Hájková, V. Altapova, T. Burian, A. J. Gleeson, L. Juha, M. Jurek, H. Sinn, M. Störmer, R. Sobierajski, K. Tiedtke, S. Toleikis, T. Tschentscher, L. Vyšín, H. Wabnitz, and J. Gaudin, Applied Physics Letters **95**, 31111 (2009).

[33]   T. Mocek, B. Rus, M. Kozlová, J. Polan, P. Homer, L. Juha, V. Hájková, and J. Chalupský,

Optics Letters **33**, 1087 (2008).

[34] K. Kolacek, J. Schmidt, J. Straus, O. Frolov, V. Prukner, R. Melich, and P. Psota, Laser and Particle Beams **34**, 11 (2016).

[35] J. Gaudin, O. Peyrusse, J. Chalupský, M. Toufarová, L. Vyšín, V. Hájková, R. Sobierajski, T. Burian, S. Dastjani-Farahani, A. Graf, M. Amati, L. Gregoratti, S. P. Hau-Riege, G. Hoffmann, L. Juha, J. Krzywinski, R. A. London, S. Moeller, H. Sinn, S. Schorb, M. Störmer, T. Tschentscher, V. Vorlíček, H. Vu, J. Bozek, and C. Bostedt, Physical Review B **86**, 24103 (2012).

## LIST OF TABLES

TABLE I. Single-shot ablation thresholds of amorphous carbon and fullerene. Error of threshold determination varies around 20 %.

| Material | Radiation source | Wavelength (nm) | Damage threshold (mJ/cm$^2$) |
|----------|------------------|-----------------|------------------------------|
| a-C | SCSS | 60 | 82 |
| $C_{60}$ | SCSS | 60 | 23 |
| a-C | FLASH | 13.5 | 384 |
| $C_{60}$ | FLASH | 13.5 | 41 |

TABLE II. Coefficients used in the van der Waals softly cut potential, Eq. (2) and (A1).

| vdW Coefficient | Value |
|-----------------|-------|
| $C_{12}$ | 22500 (eV·Å$^{12}$ ) [23] |
| $C_6$ | 15.4 (eV·Å$^6$) [15] |
| $a$ | 0.1286478847 (eV/Å$^5$) |
| $b$ | -1.858707955 (eV/Å$^4$) |
| $c$ | 10.66718892 (eV/Å$^3$) |
| $d$ | -30.40360058 (eV/Å$^2$) |
| $e$ | 43.05091787 (eV/Å) |
| $f$ | -24.23710839 (eV) |
| $a_L$ | -0.825344·10$^{-3}$ (eV/Å$^3$) |
| $b_L$ | 0.013137408 (eV/Å$^2$) |
| $c_L$ | 0.068511744 (eV/Å) |
| $d_L$ | 0.1163980800 (eV) |

## LIST OF FIGURES

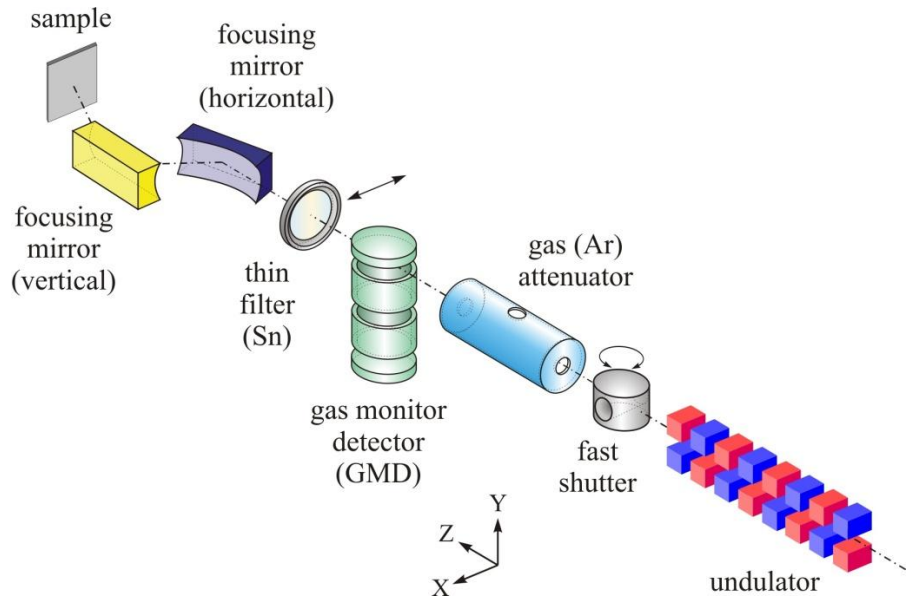

Figure 1. An experimental layout for irradiation of solid samples by focused SCSS laser beam.

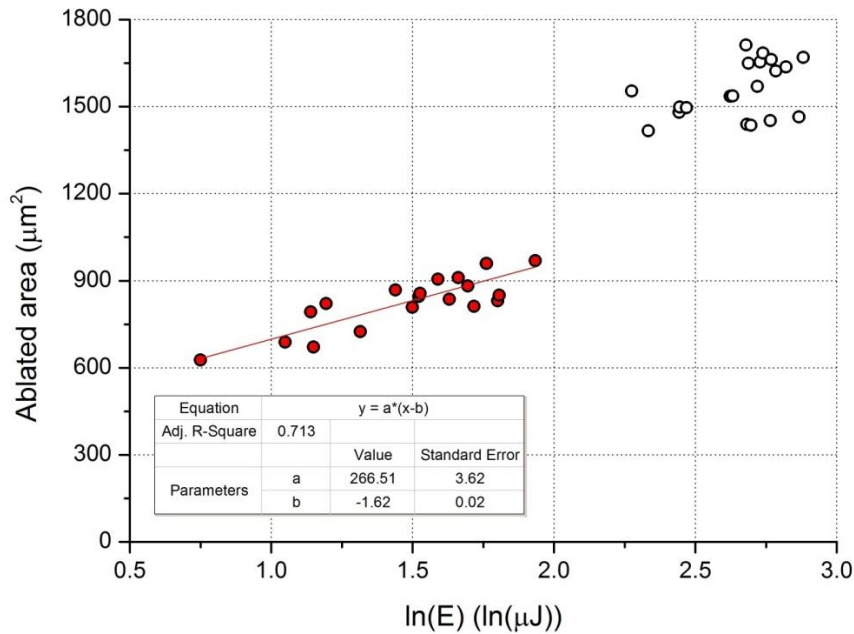

Figure 2. An example of the threshold energy determination for a-C sample irradiated with SCSS laser pulses. Low energy part (red dots) is fitted with a linear function assuming a Gaussian-like narrow central peak (fit parameters are listed in the inset). Ablation threshold energy is found to be 197.9 ± 4.6 nJ.

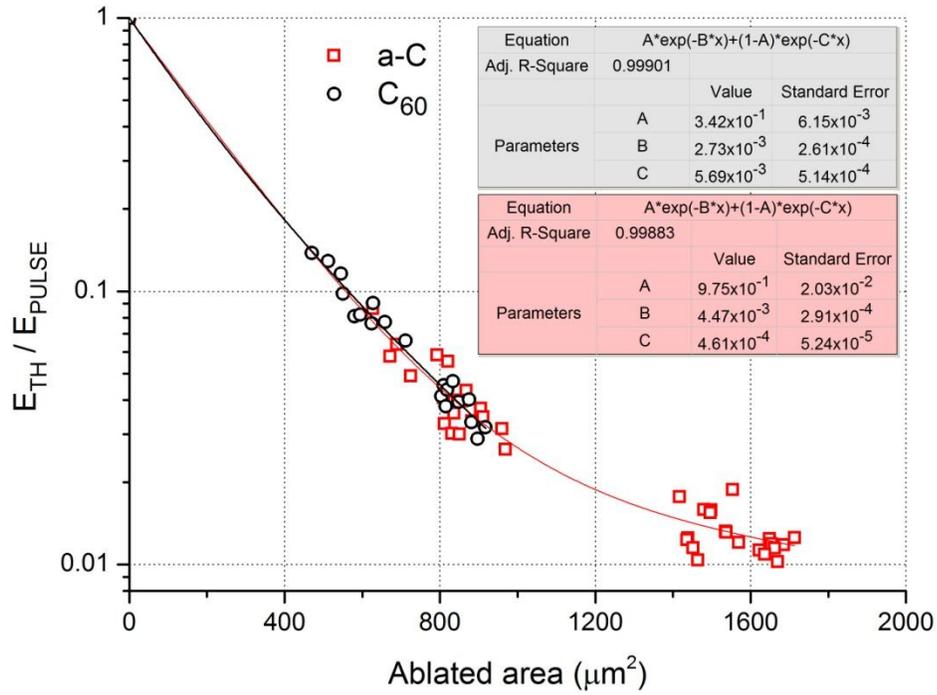

Figure 3. Comparison of f-scan plots obtained for a-C (red squares) and $C_{60}$ (black circles) samples irradiated by an SCSS single shot, interpolated by exponential functions (parameters are listed in the insets). Derived effective areas are used for fluence threshold evaluation for a particular material. Effective area values of $241 \pm 16$ µm$^2$ and $272 \pm 24$ µm$^2$ follow from the plots represented here for a-C and $C_{60}$, respectively.

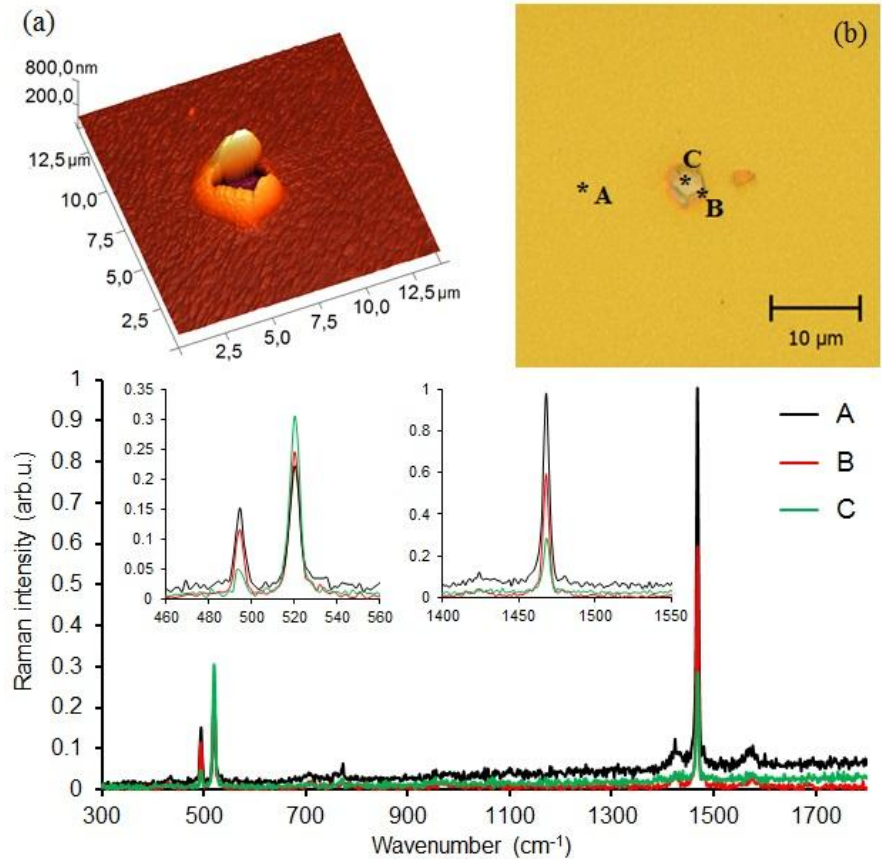

Figure 4. $C_{60}$ layer irradiated by a single ultra-short pulse of 13.5 nm FEL radiation at a fluence well above the single shot ablation threshold: (a) AFM reconstructed surface topography (two insets are zooming in on the peaks) and (b) DIC micrograph of the damage pattern with marked positions where (c) Raman spectra were acquired. A shape of the main Raman peak at 1469 cm$^{-1}$ assigned to the pentagonal pinch mode of $C_{60}$ looks unchanged inside and outside the damage pattern (see the inset on the right). The same behavior indicates a band belonging to the $A_g(1)$ mode of $C_{60}$ (can be seen at 490 cm$^{-1}$ in the inset on the left).

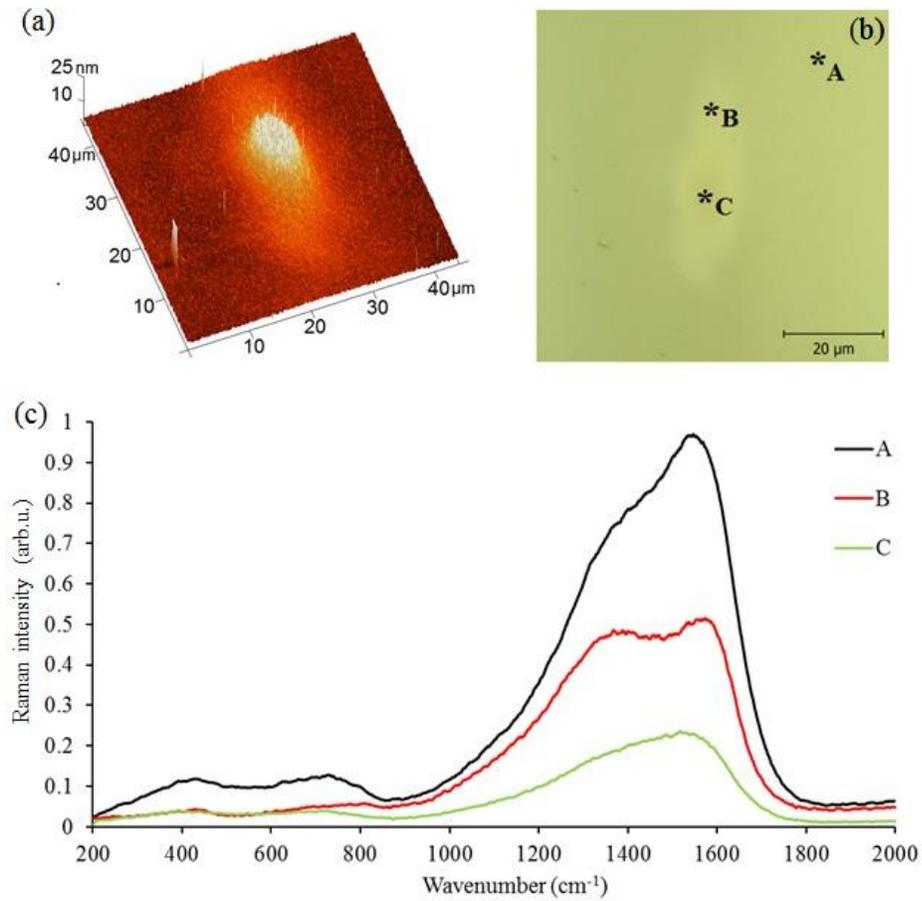

Figure 5. A layer of a-C irradiated by a single ultra-short pulse of 60 nm radiation focused above the single shot damage threshold at the SCSS facility: (a) AFM reconstructed surface topography and (b) DIC micrograph of the damage pattern with marked positions where (c) Raman spectra were acquired.

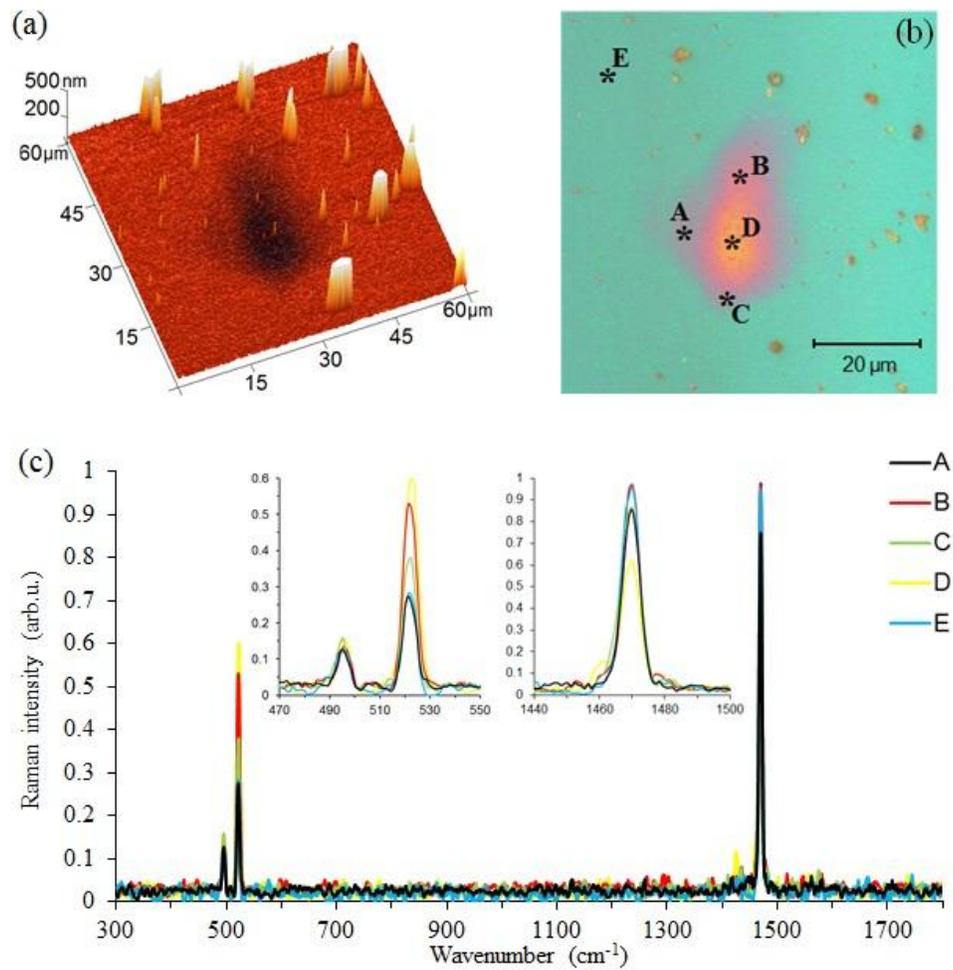

Figure 6. $C_{60}$ sample irradiated by a single ultra-short pulse of 60 nm radiation focused slightly above the single shot ablation threshold at the SCSS facility: (a) AFM reconstructed surface topography and (b) DIC micrograph of the damage pattern with marked positions where (c) Raman spectra were acquired.

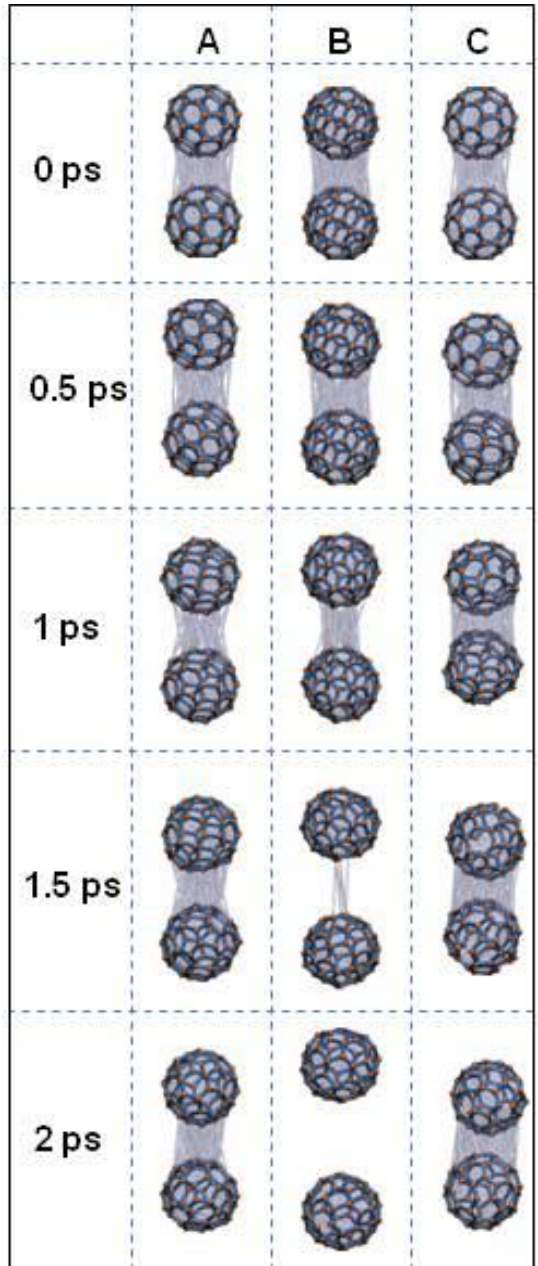

Figure 7. Snapshots of atomic positions in scc $C_{60}$ crystal at different times after the FLASH irradiation, calculated with XTANT. The following cases are compared: (A) allowing for electron emission at the under-critical dose; (B) allowing for electron emission at the over-critical dose; and (C) 'artificially' excluded electron emission (thus excluding the Coulomb explosion effect) at the 'above-threshold' absorbed dose. Orange balls represent C atoms, thick blue lines depict covalent bonds, semi-transparent grey lines depict van der Waals bonds.

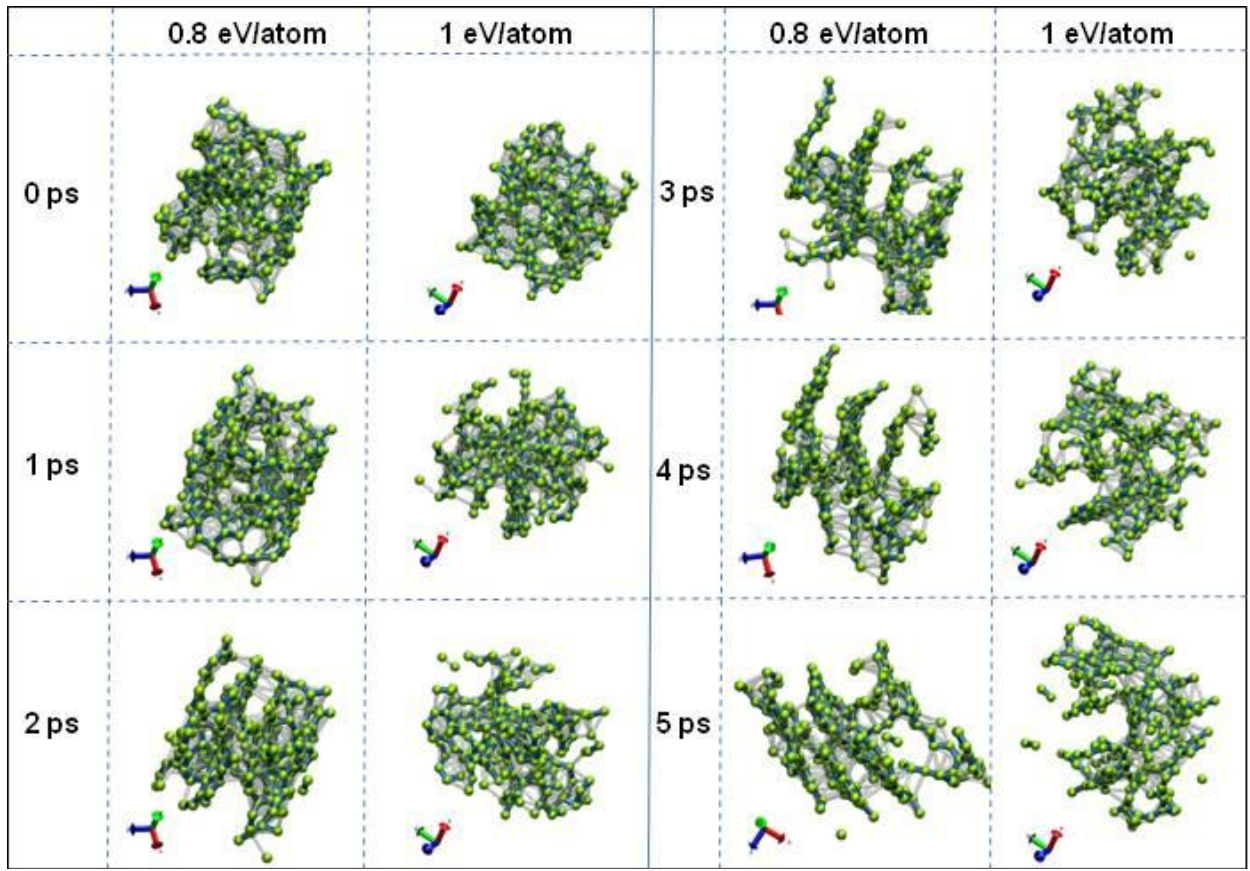

Figure 8. Calculated atomic snapshots of a-C irradiated with the FLASH pulse (92 eV photon energy, 30 fs FWHM) for the below-threshold dose (0.8 eV/atom), and the above-threshold dose of 1 eV/atom at different time instants. 216 atoms were in the modeled supercell. Green balls represent C atoms, blue lines depict covalent bonds, transparent grey lines depict van der Waals bonds (note that not all the vdW bonds are shown to keep the picture transparent).

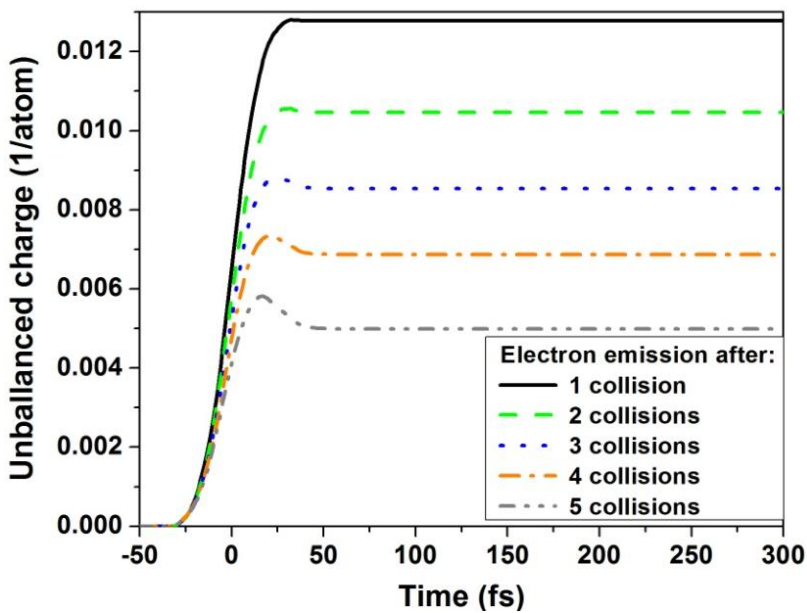

Figure 9. Build-up of the unbalanced charge in $C_{60}$ molecules due to electron escape, calculated with XTANT. Predictions obtained with various estimates of the number of collisions, after which an electron was emitted, are compared. The pulse parameters of this study case are: 92 eV photon energy, 30 fs pulse duration, and the absorbed dose of 0.6 eV/atom.

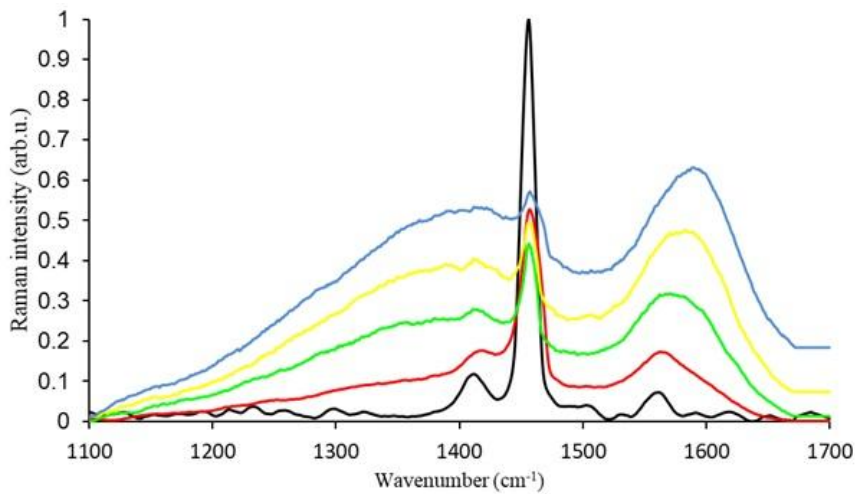

Figure 10. The Raman spectra taken in the same spot on fullerene sample excited and modified by cw $Ar^+$ laser micro-beam (514.5-nm). Black line: pristine; red: 2-min irradiation; green: 4-min irradiation; yellow: 6-min irradiation; blue: 8-min irradiation.

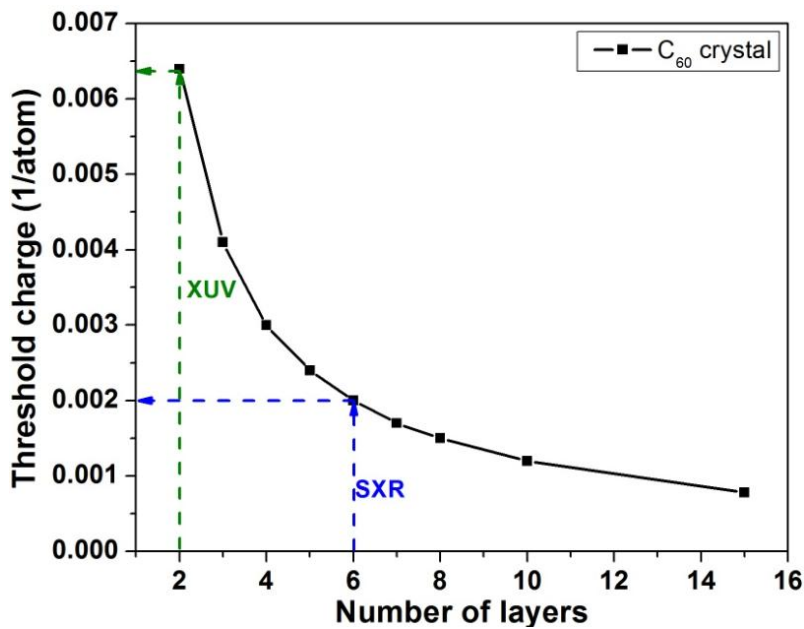

Figure 11. Estimated damage threshold charge in $C_{60}$ crystal as a function of the number of crystal layers with unbalanced charge. Arrows indicate the number of excited layers from where electrons can escape in case of SXR and XUV irradiation.

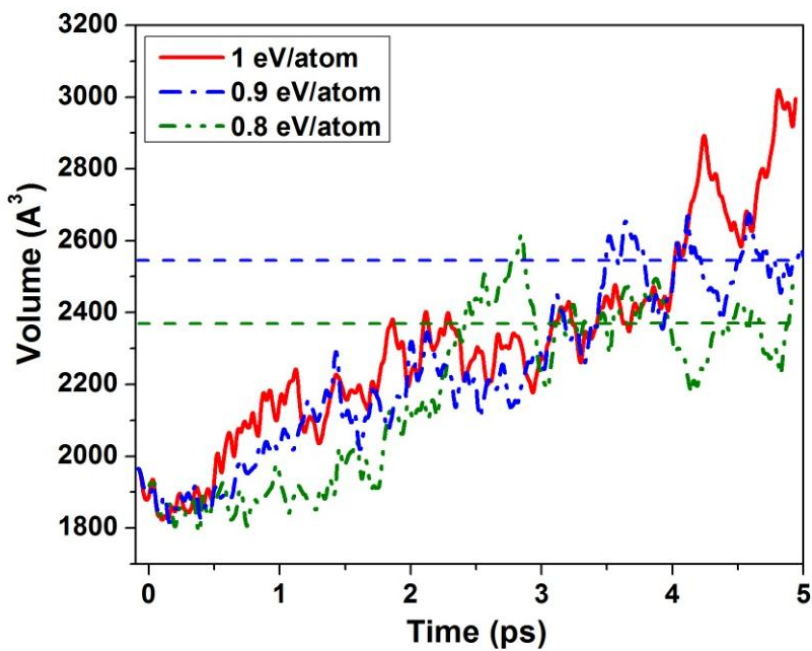

Figure 12. Transient volume of the simulated supercell (with 216 atoms) of amorphous carbon for different deposited doses, calculated with XTANT. Horizontal dashed lines indicate the saturation of the sample expansion at absorbed doses of 0.8 and 0.9 eV/atom. In the case of 1 eV/atom dose, the volume expansion continues.